\documentclass[aps,pre,showpacs,twocolumn]{revtex4}
\usepackage{amssymb}
\usepackage{graphicx}
\usepackage{colordvi}
\usepackage{color}
\usepackage{dcolumn,amsmath,amsthm,amscd,amsfonts,amssymb,epsfig,graphics,graphicx,eucal}

\newcommand{\be}{\begin{equation}}
\newcommand{\ee}{\end{equation}}

\def\phi{\varphi}

\def\bc{\begin{center}}
\def\ec{\end{center}}

\begin{document}

\title{Resonant scattering of two-dimensional honeycomb ${\cal{PT}}$ dipole structure}
\author{P. Marko\v s$^{1}$,  and  V. Kuzmiak$^{2}$}
\affiliation{%
$^1$Department of Experimental Physics, Faculty of  Mathematics Physics and Informatics, Comenius University in Bratislava, 842 28 Slovakia
\\
$^2$  Institute of Photonics and Electronics, Academy of Sciences of the Czech Republic,v.v.i., Chaberska 57, 182 51, Praha 8, Czech Republic
}

\date{\today}

\begin{abstract}
We studied numerically electromagnetic response of the finite periodic structure consisting of the ${\cal{PT}}$ dipoles represented by two infinitely long, parallel cylinders with the opposite sign of the imaginary part of a refractive index which are centered at the positions of two-dimensional honeycomb lattice. We observed that the total scattered energy reveals series of sharp resonances at which the energy increases by two orders of magnitude and an incident wave is scattered only in a few directions given by spatial symmetry of periodic  structure. We explain this behavior by analysis of the complex frequency spectra associated with an infinite honeycomb array of the ${\cal{PT}}$ dipoles and identify the lowest resonance with the broken ${\cal{PT}}$-symmetry mode formed by a doubly degenerate pair with complex conjugate eigenfrequencies corresponding to the $K$-point of the reciprocal lattice.
\end{abstract}
\pacs{42.70.Qs}
\maketitle

\input{epsf.tex}
\epsfverbosetrue

\section{Introduction}
\label{intro}

New materials with unexpected electromagnetic(EM) response are highly desirable for technological applications and, likewise, for our understanding of electromagnetic properties of matter. In particular, ${\cal{PT}}$ symmetric optical systems based on the $\cal{PT}$ concept \cite{bender1,bender2} which was proposed in quantum mechanics as an alternative criterion for non-Hermitian Hamiltonians that possesses a real eigenspectrum, attracted great deal of attention recently. They exhibit unconventional behavior and enable novel capabilities for control of light propagation\cite{iop}-\cite{wimmer}.  Several types of extended systems included $\cal{PT}$ gratings, $\cal{PT}$ lattices and $\cal{PT}$-symmetric resonant structures characterized by the complex-valued periodic functions have been studied both theoretically and experimentally\cite{feng2}-\cite{phang2}. The periodic systems also provide asymmetric response, offering for example a unidirectional invisibility \cite{lin}, unidirectional transmission\cite{peng}, unidirectional lasing\cite{lee2} and unidirectional energy transfer in both linear\cite{guo},\cite{ruter,longhi3} and nonlinear systems\cite{mussli},\cite{ramez,vvk}.

Majority of the previous works have focused on either effective zero-dimensional systems, such as coupled cavity resonators or one-dimensional systems such as waveguide arrays\cite{regen,feng1}, ring lasers \cite{feng2}, and one-dimensional lattice systems \cite{prl17} that have been shown to possess many interesting features associated with ${\cal{PT}}$ symmetry.
A straightforward extension to two dimensions does not necessarily leads to novel effects but requires implementation of nontrivial $\cal{PT}$ symmetry which may introduce 2D peculiarities. For example, introducing a nonfactorizable unidirectional coupling between the plane wave components was shown to give rise to a nonreciprocal chirality and asymmetric transmission with respect to the direction of propagation of an incident wave\cite{kestas1}. Another interesting concept which may provide additional flexibility in tailoring flow of light is based on the local $\cal{PT}$ symmetry which gives rise to nontrivial flows such as axisymmetric \cite{kestas2} or any other flow configuration on demand \cite{kestas3}.
A systematic exploration of higher dimensional ${\cal{PT}}$-symmetric systems \cite{ge}-\cite{cerjan} is associated with recent discovery of thresholdless ${\cal{PT}}$ transitions, for which the ${\cal{PT}}$ symmetry is spontaneously broken in the presence of infinitesimal amount of gain and loss\cite{ge}.
General conditions for thresholdless ${\cal{PT}}$ symmetry were proved in both 2D and 3D systems\cite{ge}. Realization of the thresholdless ${\cal{PT}}$ symmetry
transitions is of interest as they reduce the experimental requirements for observing exceptional point physics. In addition, such a system enables a new form of band structure
engineering, and can result in various effects(e.g. ${\cal{PT}}$-superprism effect\cite{cerjan}) which are distinct from those observed in Hermitian photonic crystals.

Recently, ${\cal{PT}}$ dipole was proposed as a building block for construction of finite periodic structures mimicking the macroscopic
${\cal{PT}}$-materials and its electromagnetic response has been studied in detail in \cite{smk}. Owing to their spatial asymmetry and the presence of both the gain/loss elements, ${\cal{PT}}$ combination of dipole promises to exhibit an unusual EM response, not observable in dielectric structures.
Indeed, such 2D complex structures built from the individual molecules may possess richer variety of the peculiar effects associated with ${\cal{PT}}$-symmetry and arising from higher dimensionality provided by an additional degree of freedom of such structure.

In this paper we analyze the electromagnetic response of finite structure composed from ${\cal{PT}}$ dipoles represented by gain-loss cylinders arranged in a 2D honeycomb lattice.
We calculate the radiated field and show that in the far field limit the system radiates the energy only in a few directions given by the spatial symmetry of the structure. Such a behavior can be understood in the context of 2D photonic lattices which are known to provide an ability to control of EM field that originates from their unique properties. For example, it has been demonstrated that lasing in photonic crystal based laser structures\cite{noda} occurs at specific points on the Brillouin zone boundary and at points of band crossing and splitting when optical gain is supplied. At these points, or band edges, waves propagating in different directions couple, and a standing wave is formed. We will show that scattering properties of the
2D ${\cal{PT}}$-symmetric lattice reflect the peculiarities of the complex band structure associated with a specific distribution of gain-loss cylinders and, in particular, that the
sharp resonances can be associated with the ${\cal{PT}}$-broken mode corresponding to high symmetry points in the first Brillouin zone.

Our paper is organized as follows. In Sec. II we describe the ${\cal{PT}}$ dipole structure considered in numerical experiment carried out by using algorithm based on the expansion of the electromagnetic field into cylinder functions and define the relevant scattering characteristics. In Sec. III present numerical results obtained for the total scattering energy as a function of the normalized frequency and scattering diagrams associated with the resonances observed in the spectra. We explain these phenomena in terms of the complex band structure associated with an infinite 2D honeycomb lattice presented in Sec. IV. and discuss the scaling of the total scattered energy with the size of the sample.

\def\sr{S_{\textrm{\footnotesize{Re}}}}
\def\si{S_{\textrm{\footnotesize{Im}}}}

\section{Structure and Scattering}

\begin{figure}[t!]
\bc
\includegraphics[width=0.35\textwidth]{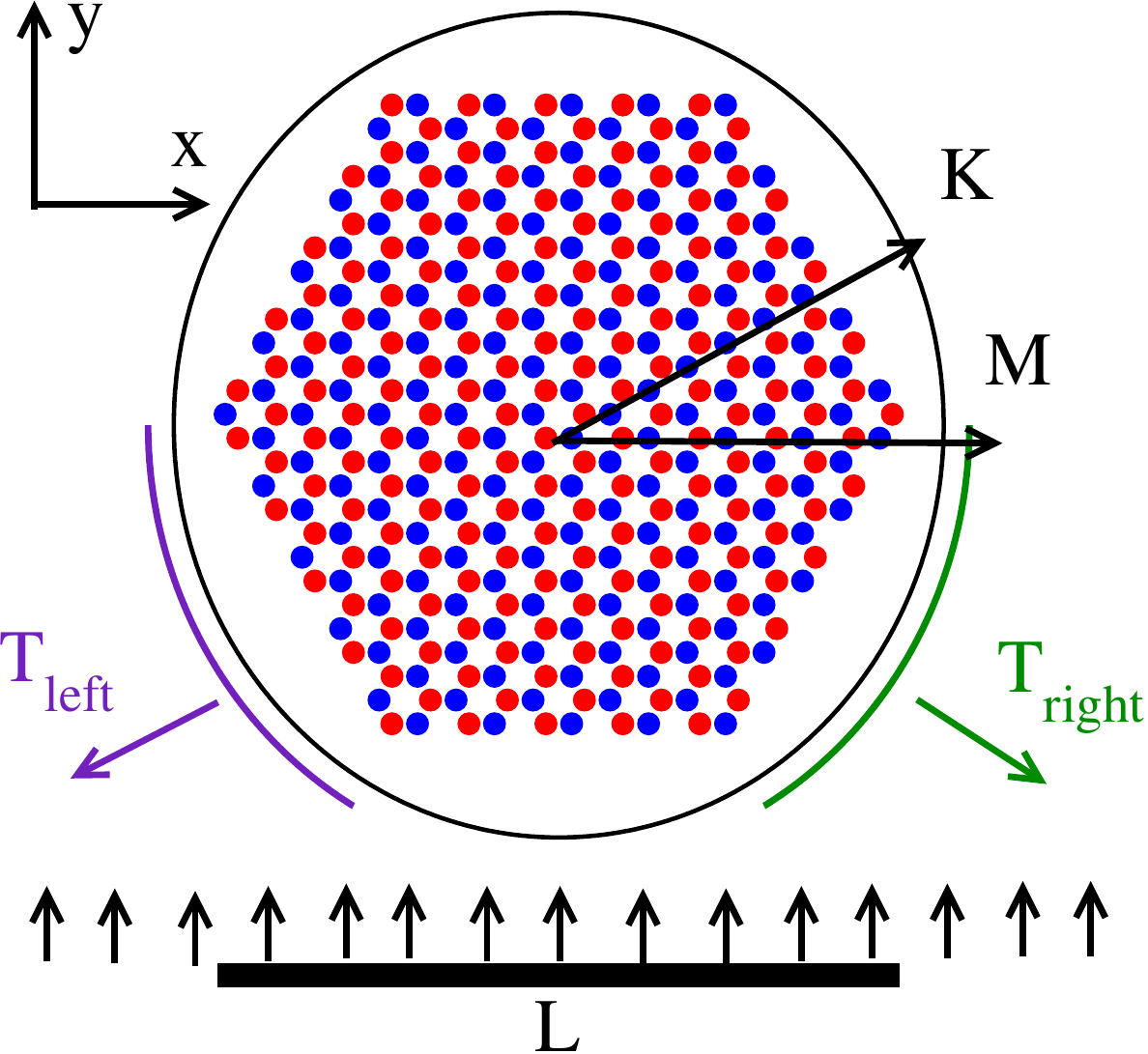}
\ec
\caption{(Color online) The ${\cal{PT}}$ dipole structure considered in  the scattering experiment.  Red and blue circles represent cylinders
 with gain and loss, respectively. Incident plane wave propagates along the $y$ direction, $L$ determines the maximal size of the structure in the $x$ direction (geometrical shadow). Scattered energy is calculated along the circumference of the circle centered at the center of mass of the structure.
The black arrows indicate two directions of propagation along the $\Gamma K$ and $\Gamma M$ directions defined in the reciprocal lattice. $T_{\rm left}$ and $T_{\rm right}$ refer to a (normalized) portion of the energy radiated through displayed  $\pi/3$-segments of the circle.}
\label{sh-1}
\end{figure}

The structure under consideration which is shown in Fig. \ref{sh-1} represents a finite size photonic honeycomb lattice where spatially alternating gain/loss is added. Specifically, it consists of $N=183$  ${\cal{PT}}$ dipoles each of them formed by two gain/loss infinitely long cylinders, parallel to the $z$-axis.  The lattice constant is determined by the distance between centers of the cylinders which defines the size of the dipole, $a=1$. The gain/loss cylinders with radius $R_0 = 0.45a$ are characterized by the refractive index  $n= 1.1 \pm 0.1 i$. We note that the corresponding infinite structure is ${\cal{PT}}$ symmetric along $\Gamma M$ and two additional equivalent directions obtained by rotation by $\pi/3$ and $2\pi/3$ angles belonging to threefold symmetry of the lattice.

The incident EM plane wave which is assumed to propagate in the $xy$ plane along the $y$ direction with unit amplitude and normalized frequency $f = a/\lambda$
\be
E_{\rm inc} \equiv E_z(x, y |\omega)_{\rm inc} =  \exp[i k_y y - i\omega t]
\label{eq.P1}
\ee
is polarized parallel to the axes of the cylinders.
The scattered energy is calculated on the circumference indicated in Fig. \ref{sh-1} by the black circle
with radius $R\gg L$ centered at the center of mass of the structure.
Specifically, we evaluate the radial component of the Poynting vector
$P(R,\phi) = E_z(R,\phi)[H_t(R,\phi)]^{*}$ where $H_t(R,\phi)$ is
the tangential component of the magnetic field and determine the scattering variable
\be\label{eq:s}
S(R) = R\int_0^{2\pi} P(R,\phi) d\phi
\ee
which represents the total energy radiated from the structure.
We calculate the scattering factor
\be
Q= S/S_0
\label{eq:Q}
\ee
where $S_0 = E_{\rm inc}H^*_{\rm inc}L$ is the energy incident to the structure \cite{Hulst}.
We also define the quantities
\be
T_{\rm right} = \frac{R}{S}\int_{5/3\pi}^{2\pi} P(R,\phi) d\phi
\label{eq.tl}
\ee
and
\be
T_{\rm left} = \frac{R}{S}\int_{\pi}^{4/3\pi} P(R,\phi) d\phi,
\label{eq.tr}
\ee
which characterize the right/left symmetry of the electromagnetic response.
To compute the scattering characteristics defined above we apply a numerical algorithm based on the expansion of electromagnetic field into
cylinder functions \cite{Hulst}. Our approach is described in detail in Ref.~\onlinecite{PM3}.

\begin{figure}[b!]
\bc
\includegraphics[width=0.3\textwidth]{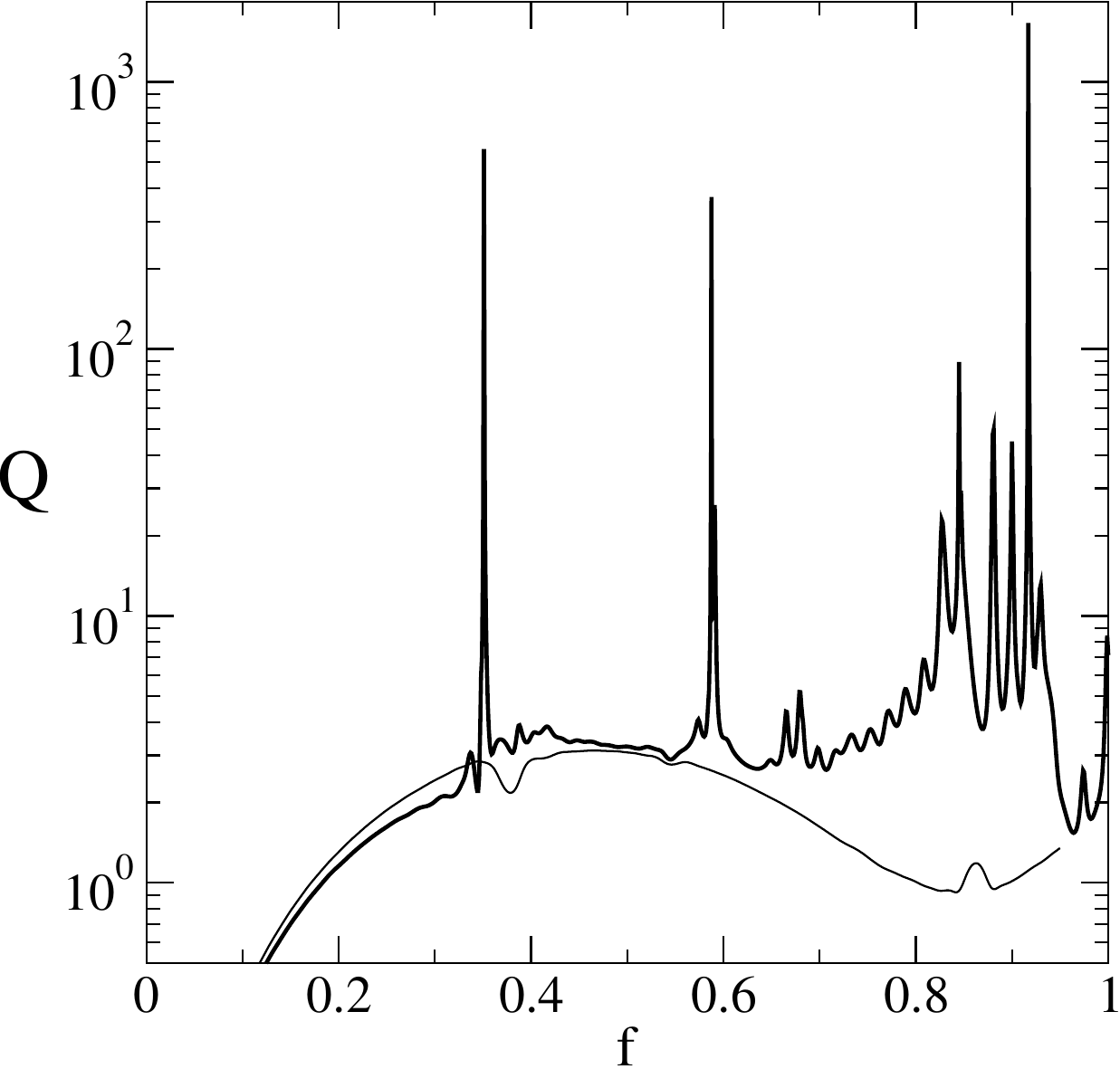}
\ec
\caption{%
Scattering factor $Q$ of the structure shown in Fig. \ref{sh-1} as a function of the frequency of incident plane wave.
Thin line displays $Q$ for  the same structure except the cylinders have real refractive index $n=1.1$.
}
\label{S}
\end{figure}

\section{Resonances in a finite 2D honeycomb lattice}

In this section we present the numerical results which demonstrate a profound effect on the scattering properties of the photonic structures when alternating gain/loss is added. In Figure \ref{S} we show the frequency dependence of the resonance factor $Q$ (normalized scattered energy)  given by Eq. \ref{eq:Q},  associated with the structure shown in Fig. \ref{sh-1} induced by an incident plane wave with unit amplitude. The total scattered energy $S$ associated with gain/loss cylinders exhibits series of sharp resonances at which scattered energy  increases by two orders of magnitude. Obtained data are compared with that for the sample of the same geometry consisting of dielectric cylinders indicated by a thin line in Fig. \ref{sh-1} which
reveals slowly varying behavior.

\begin{figure}[t!]
\bc
\includegraphics[width=0.2\textwidth]{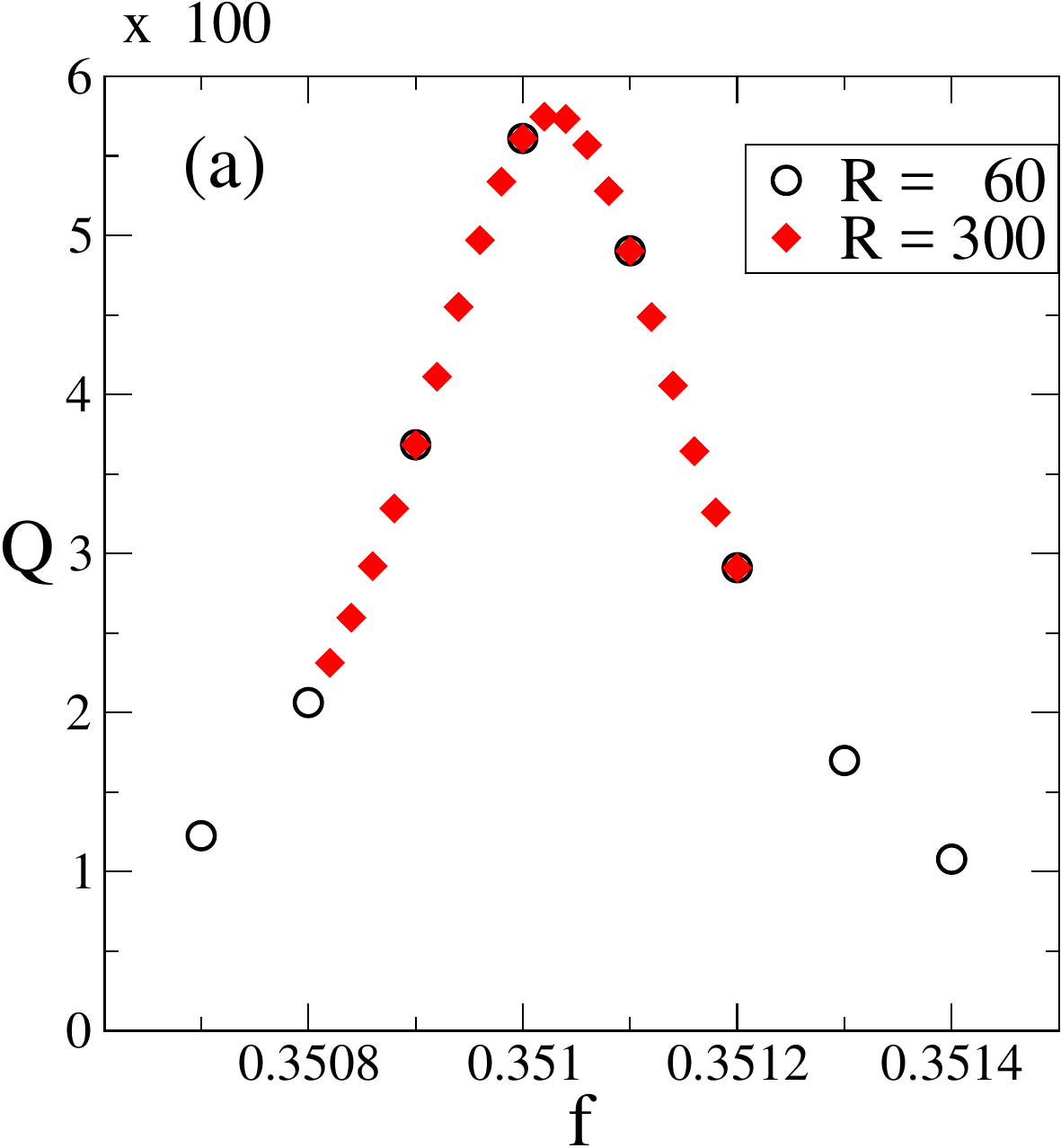}
~~~~
\includegraphics[width=0.23\textwidth]{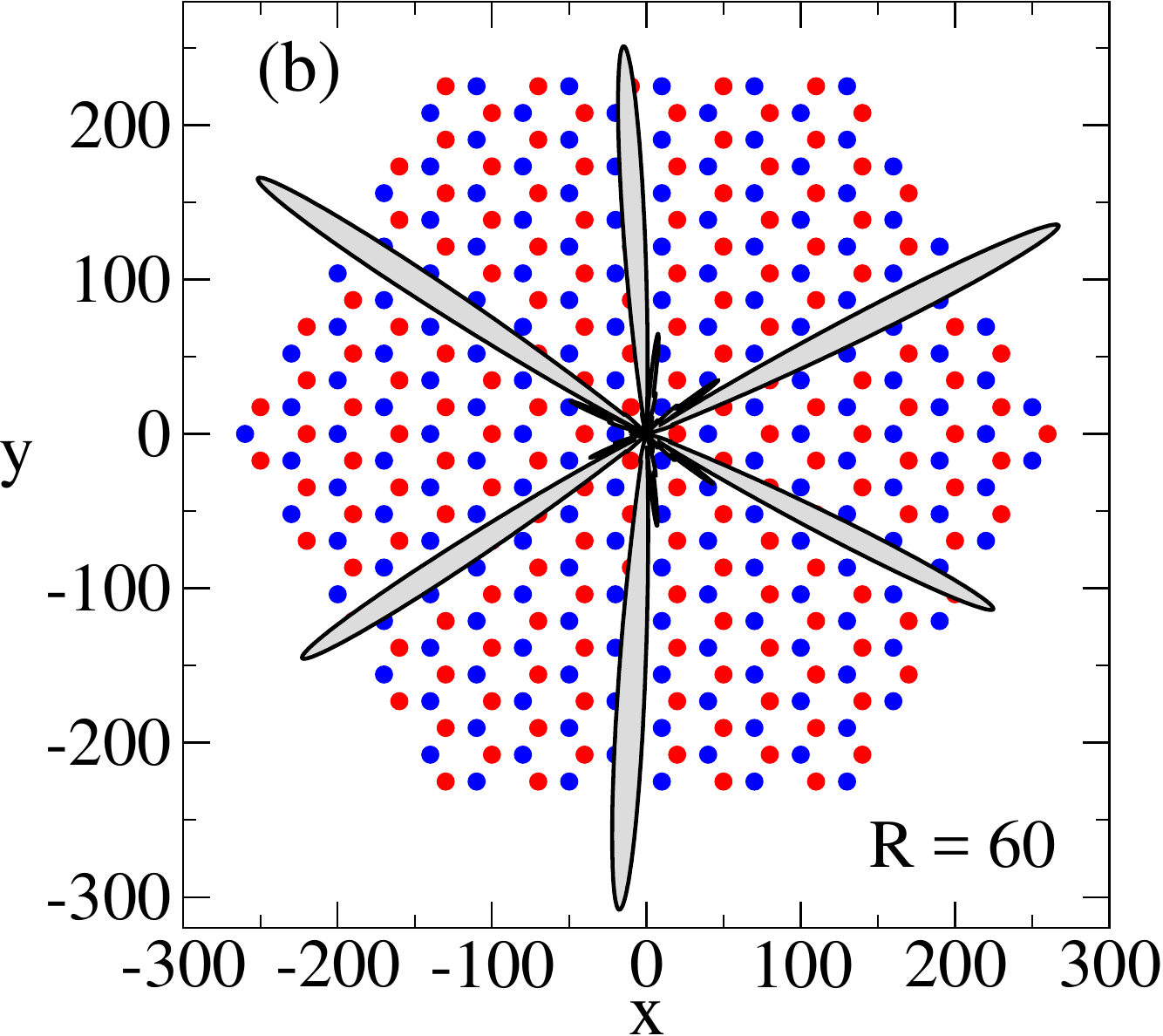}
\ec
\caption{(Color online)
(a)  Scattering factor $Q$  as a function of frequency in the vicinity of the first resonance
calculated along the circles with radii $R=60$ and $R=300$.
(b)  Scattering diagram corresponding to the radius $R=60$ exhibits
a highly-directional pattern with well defined symmetry - the incident wave is scattered along
the three equivalent $\Gamma K$ directions only.
}
\label{S-1}
\end{figure}

To obtain an additional insight into such  a striking difference we first in detail depict in Fig. \ref{S-1}a the scattering factor $Q$ in the vicinity of the first resonance $f = 0.351$. In Figure \ref{S-1}b we present the scattering diagram given by $(P(R,\phi)\cos\phi, P(R,\phi)\sin\phi)$ evaluated for the radius $R = 60$. One can see that the outgoing field is concentrated into six narrow rays oriented along the directions very close to three equivalent $\Gamma$K directions. The field scattered to other directions has much smaller intensity and is not clearly visible in Fig. \ref{S-1}b . We note that fully symmetric scattering pattern reflects the arrangement of the cylinders consisting the linear chains of the identical cylinders with gain or loss along the $\Gamma$K direction –- see Fig. \ref{sh-1}. Both the origin of the enhancement of the scattered energy and symmetry of the scattered field will be discussed in more detail in the Sec. IV.

\begin{figure}[h!]
\bc
\includegraphics[width=0.14\textwidth]{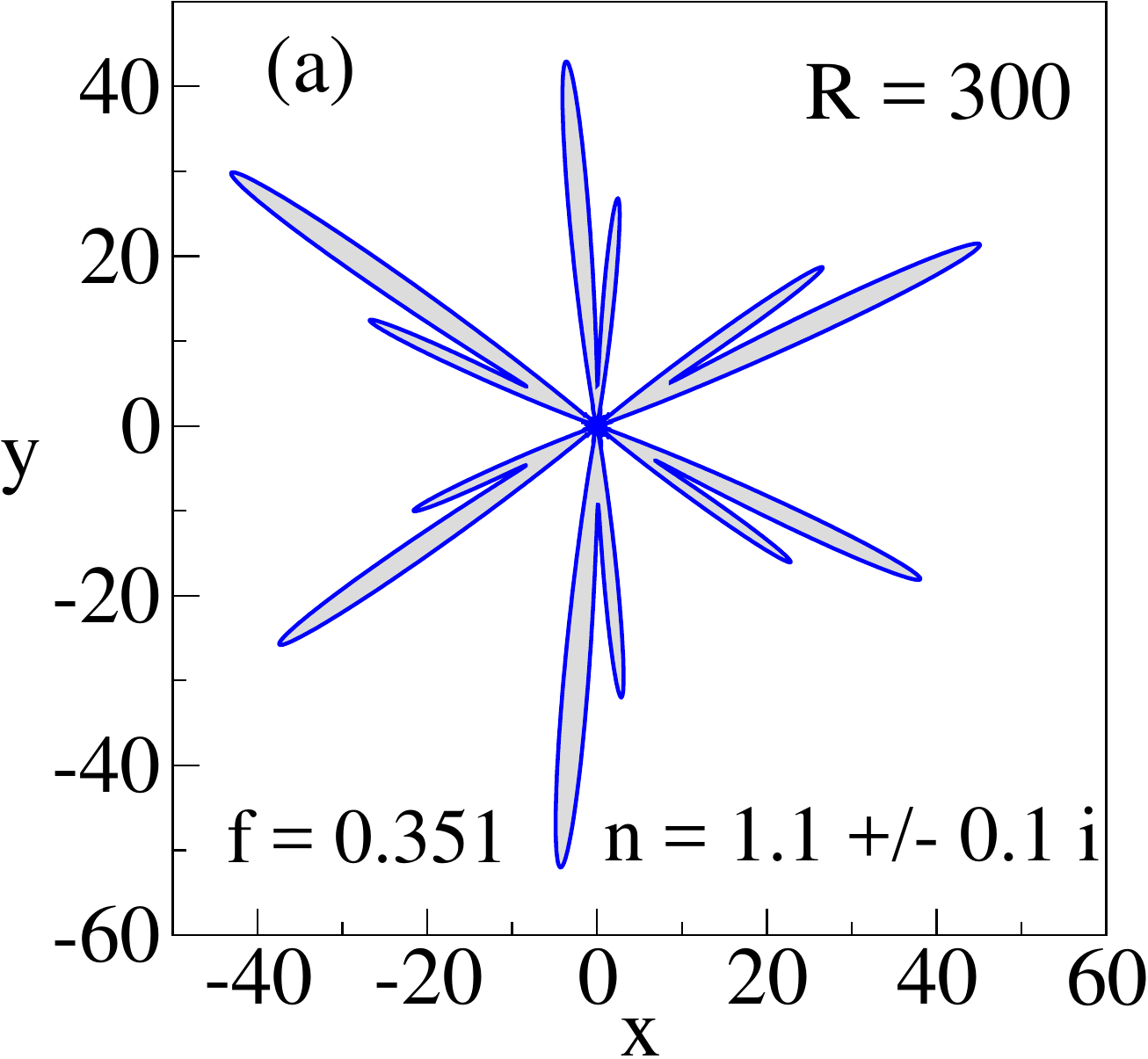}
~~
\includegraphics[width=0.14\textwidth]{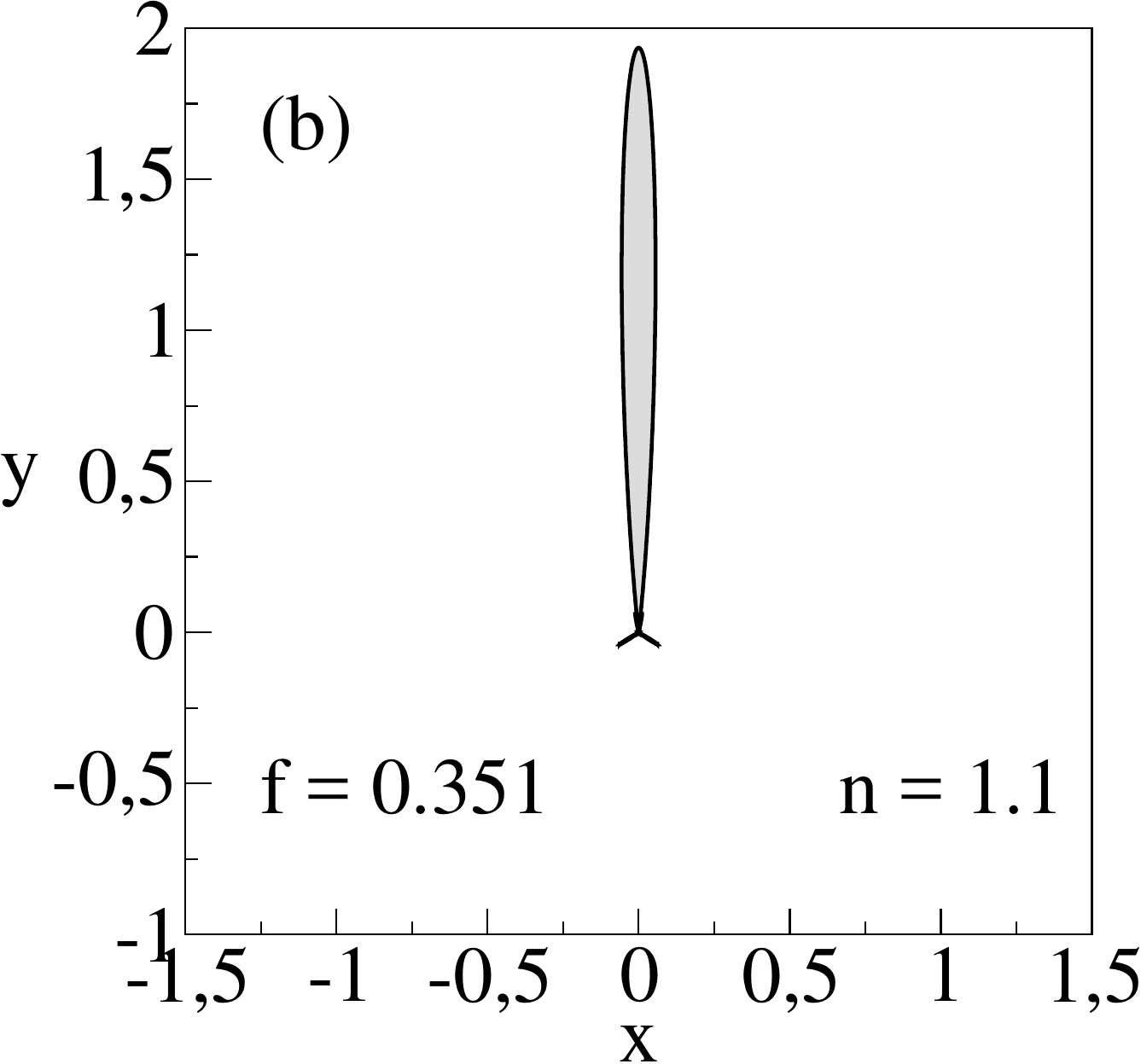}
~~
\includegraphics[width=0.14\textwidth]{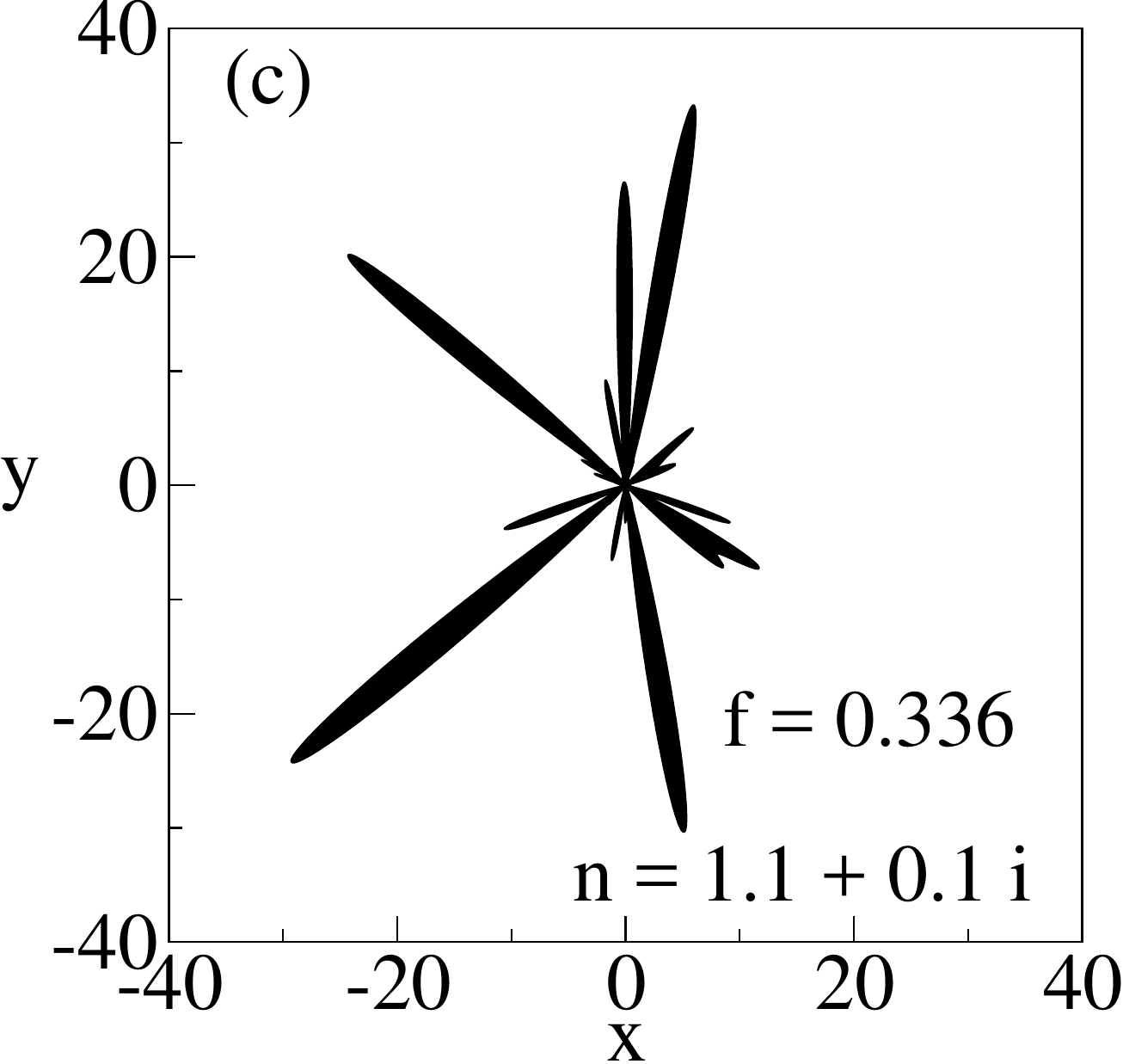}
\ec
\caption{(Color online)
Scattering diagram for  the frequency $f=0.351$ calculated along the circle with radius $R=300$. Note the splitting of each ray into two parts. That with larger intensity inclines towards the cylinder with gain (red circle in Fig. \ref{sh-1}). (b) The same albeit for dielectric cylinders. Note that the scattered energy is much smaller and scattering diagram
possesses mirror symmetry along the $y$-axis. (c) The same for structure made only from  cylinders
with gain (lossy cylinders were removed).
}
\label{S-1a}
\end{figure}

The scattering diagram evaluated at larger distance from the sample, shown in Fig. \ref{S-1a}a, reveals the existence of additional lobes with smaller amplitude. We suppose that the splitting of each ray is due to the local asymmetry of the structure (the size of the dipole is not negligible in comparison with the size of the structure). To compare scattering efficiency of ${\cal{PT}}$ dipole structure with dielectric one, we plot in Fig.\ref{S-1a}b the scattering diagram of dielectric structure with refractive index $n=1.1$. The latter result
clearly demonstrates that the incident wave undergoes only weak scattering and is consistent with a response expected for conventional periodic photonic structures characterized by a small refractive index contrast.
Figure \ref{S-1a} shows also the scattering diagram for structure from which we removed the lossy cylinders. Resulting triangular structure made from active cylinders exhibits resonant maximum at frequency $f=0.336$ with smaller amount of scattered energy $Q$ in comparison with the ${\cal{PT}}$ dipole structure and a asymmetric distribution of the scattered energy (Fig. \ref{S-1a}(c)).

\begin{figure}[t!]
\bc
\includegraphics[width=0.22\textwidth]{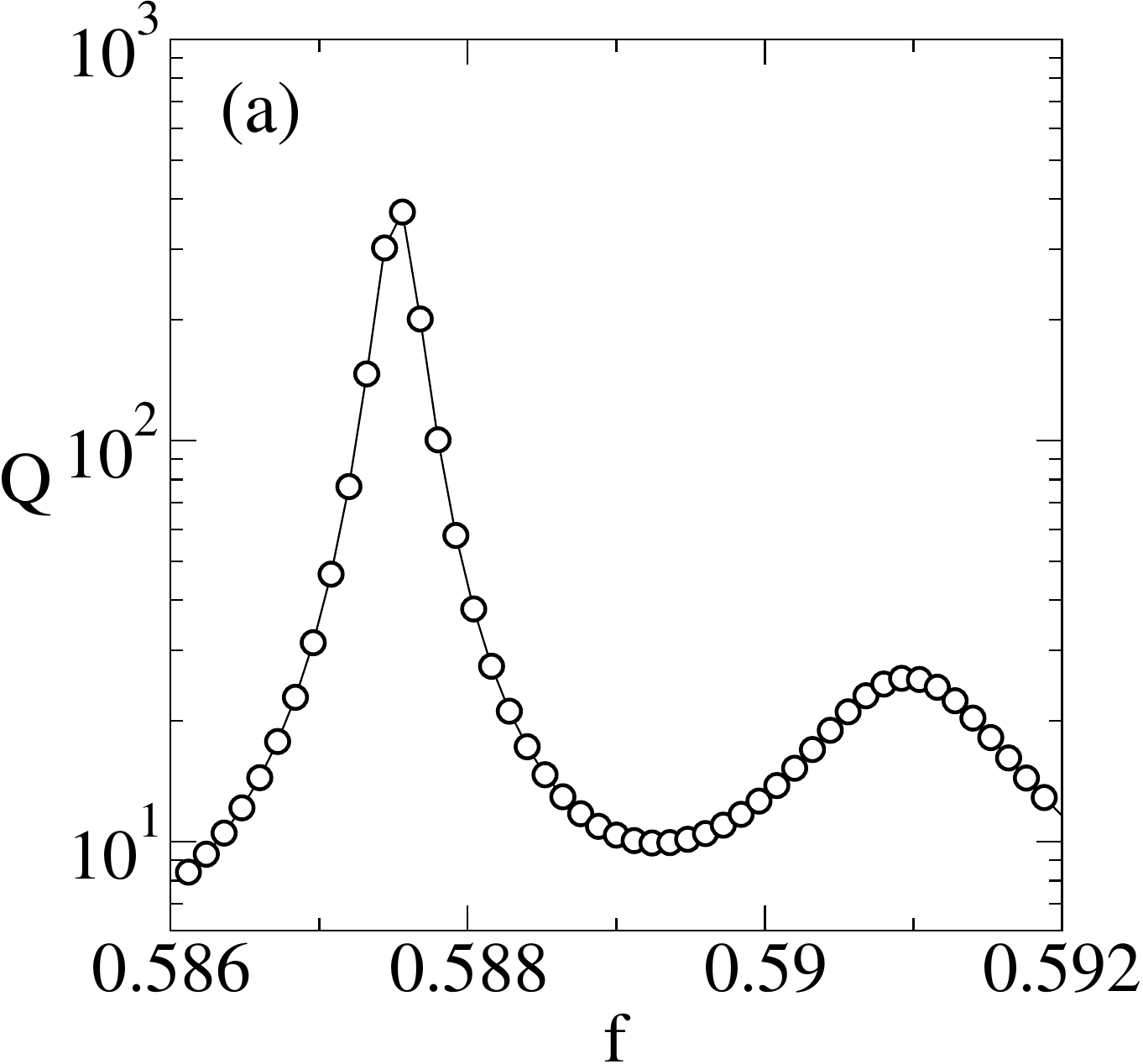}
~~~
\includegraphics[width=0.22\textwidth]{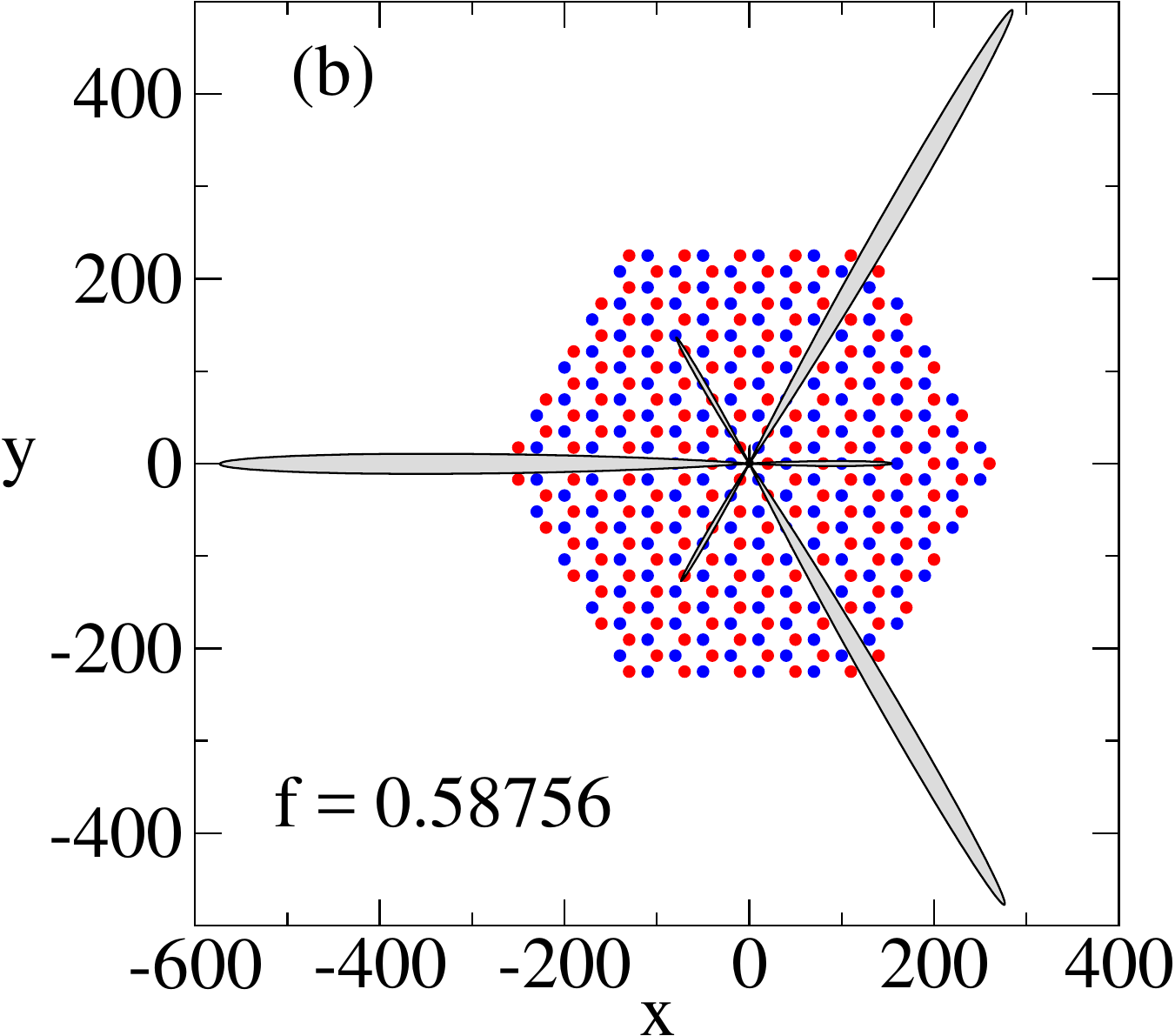}
\ec
\caption{(Color online)
(a) $Q$ as a function of the frequency for in the vicinity of the second resonance in Fig.2.
(b) Scattering diagram for the left maximum, $f=0.58756$. Note that field is scattered along $\Gamma M$ directions. Scattering diagram for the right resonance, $f=0.591$, is shown in Fig. \ref{rot}.
}
\label{S-2}
\end{figure}

\begin{figure}[b!]
\bc
\includegraphics[width=0.3\textwidth]{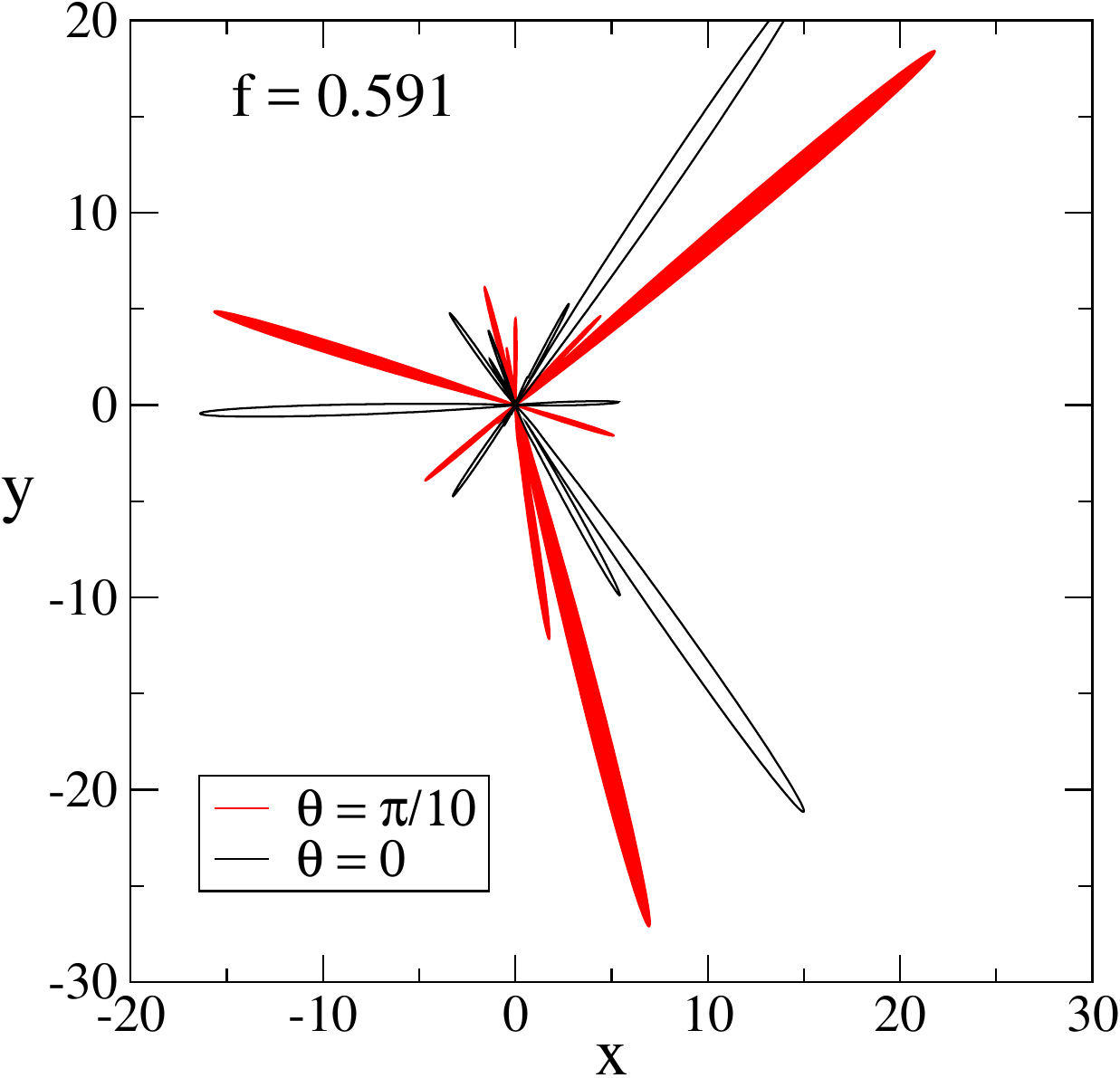}
\ec
\caption{(Color online)
Comparison of the scattering diagram for the structure rotated in $\theta=\pi/10$  with original scattering diagram (black)
confirms that the  scattering only marginally depends on the incident angle and that the mechanism of scattering is given by the sample itself.
}
\label{rot}
\end{figure}

The normalized scattered energy $Q$ and scattering diagram associated with the second lowest resonance at the frequency $f = 0.58756$ is shown in Fig. \ref{S-2}. We  observe an additional
resonance with smaller amplitude at the frequency $f = 0.591$, while the energy associated with both resonances is radiated along the $\Gamma$M directions -- see Figs. \ref{S-2}b and \ref{rot}. In contrast to the lowest resonance at $f = 0.351$, the scattering diagram exhibits a strong asymmetry which reflects the arrangement of the cylinders along $\Gamma$M direction with alternating gain and loss – see Fig.\ref{sh-1} - which is $\cal{PT}$-symmetric along the $x$-axis.  In fact, the structure along the $\Gamma$M direction consists of linear chains of $\cal{PT}$ dipoles which have been shown to give rise to asymmetric scattering \cite{smk}. This behavior will be discussed in more detail in the Sec. IV.

We also verified that the scattering diagram remains the same when the angle of incidence of the EM plane wave is changed. As an example,
we show in Fig. \ref{rot} the scattering diagram at the second resonance ($f=0.591$) calculated for the sample
rotated in the angle $\theta=\pi/10$. In spite of non-zero incident angle, the field is radiated along the same direction with respect
to the sample and confirms that mechanism of scattering is determined solely by the structure itself.

We analyzed scattering diagrams also for the frequencies outside the resonances and we found (data not shown) that the concentration of the scattered field into a few major directions is a typical property of our structure given by its spatial symmetry.

\begin{figure}[t]
\bc
\includegraphics[width=0.3\textwidth]{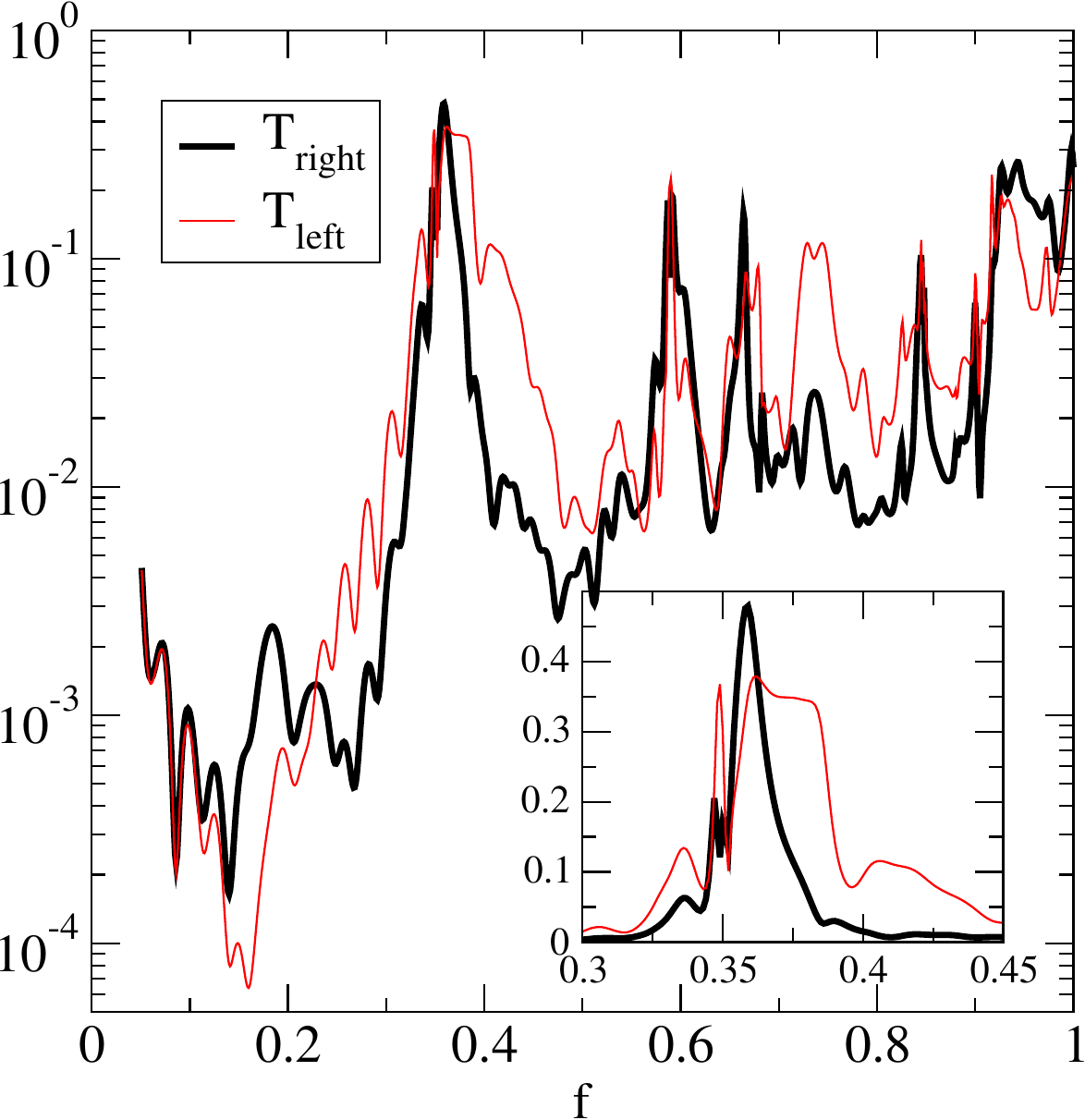}
\ec
\caption{(Color online)
Parameters $T_{\rm left}$ and $T_{\rm right}$ given by Eqs. \ref{eq.tl} and \ref{eq.tr}, respectively, as a function of frequency. Inset shows detail of the frequency dependence in the vicinity of the lowest resonance.
}
\label{par-2a}
\end{figure}

For completeness, we also show asymmetry of the scattered field induced by the introducing imaginary part in the gain-loss cylinders (Fig. \ref{par-2a}). Our data agree well with those
 calculated by the FDTD technique \cite{kestas2}.
The asymmetry of the scattering is clearly visible also from the scattering diagrams
 associated with resonant  frequencies presented above. Similar asymmetry (although without resonant effect) were observed also for dielectric structure with alternating refractive indices (1.1 and 1.2) in neighboring cylinders (data not shown).

\section{Band structure of a infinite 2D honeycomb lattice}

\begin{figure}[h!]
\bc
\includegraphics[width=0.35\textwidth]{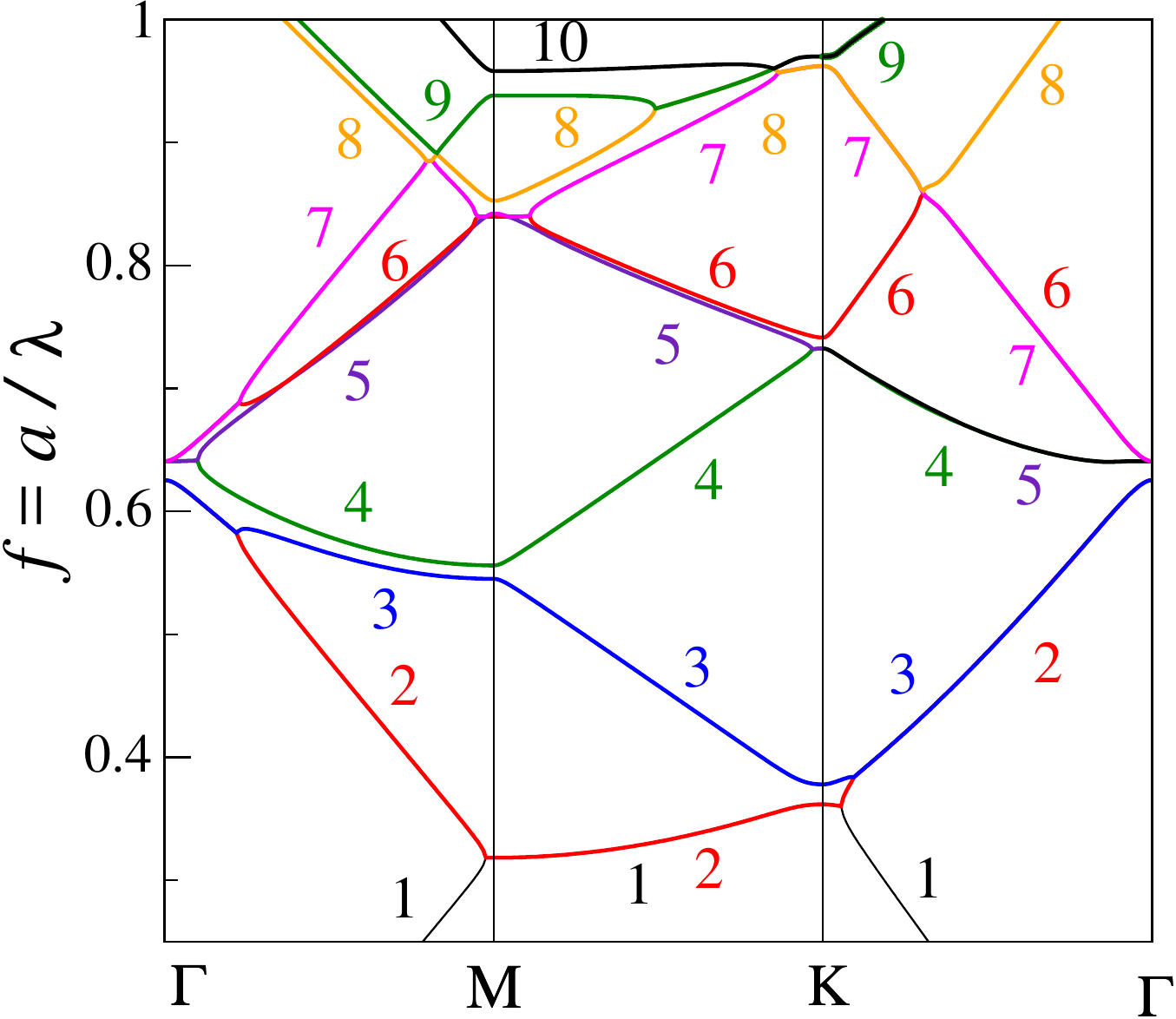}
\ec
\caption{(Color online)
Complex band structure of an infinite 2D honeycomb lattice: real part of the frequency, $n=1.1\pm 0.1i$.
Note the degeneracy of the real part of the frequency for various bands.
The degenerate bands consist of two ${\cal{PT}}$-symmetric eigenstates with positive and
negative imaginary part of the frequency shown in Fig. \ref{band-imag}.
}
\label{band-real}
\end{figure}

\begin{figure}[h!]
\bc
\includegraphics[width=0.35\textwidth]{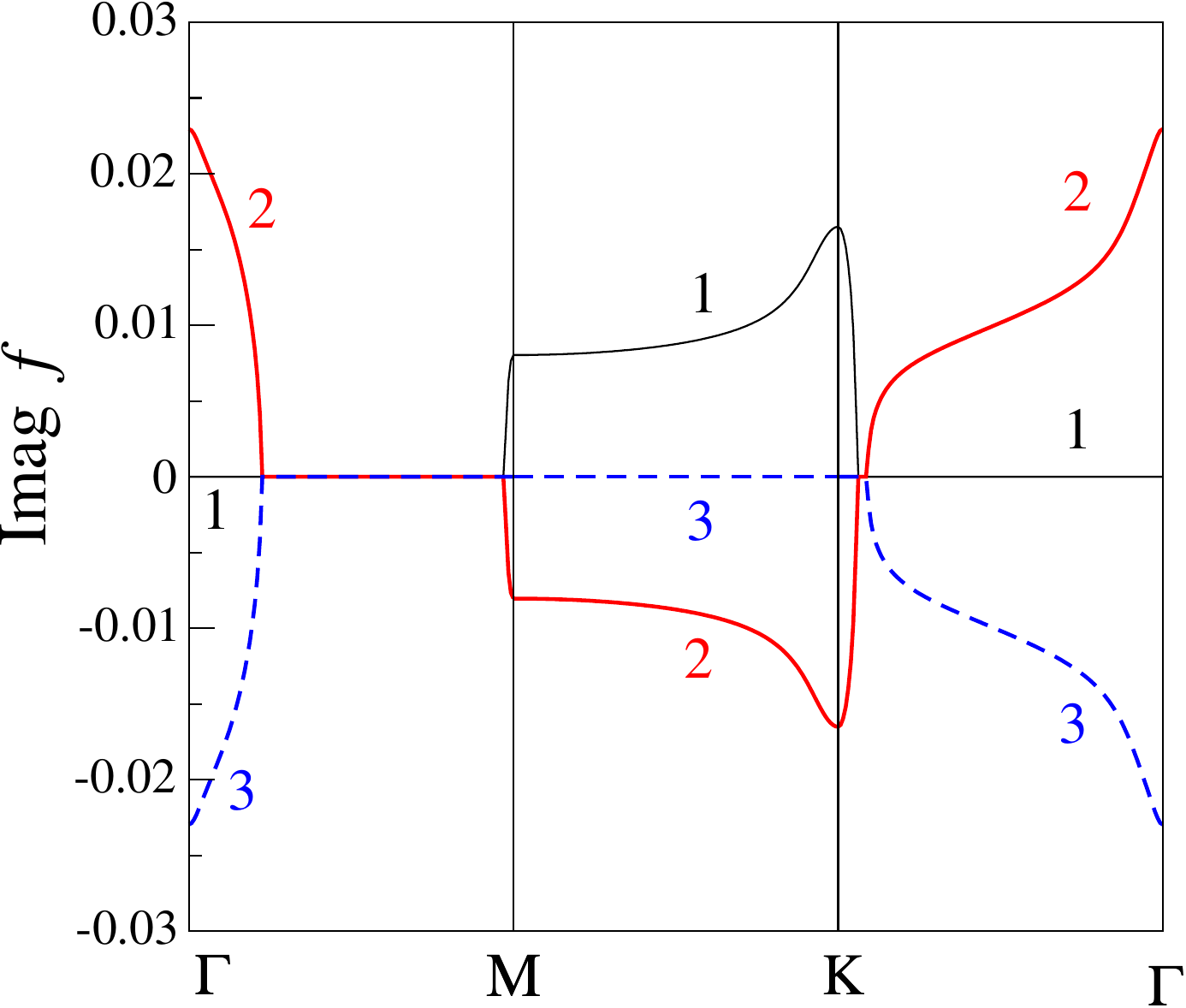}
\ec
\caption{(Color online)
Complex band structure of an infinite 2D honeycomb lattice: imaginary part of the frequency, $n=1.1\pm 0.1i$.
Only three lowest bands are shown.
}
\label{band-imag}
\end{figure}

In this section we demonstrate that the scattering diagrams shown in Figs. \ref{S-1a}-\ref{rot} can be explained in terms of the band structure of
an infinite 2D honeycomb lattice consisting of gain/loss cylinders. The band structure was calculated by using the plane-wave expansion method \cite{plihal,sakoda}.
In Figs. \ref{band-real}-\ref{band-imag} we present real and imaginary parts of the complex photonic band structure. In analogy with previous results
\cite{mock}-\cite{cerjan} on 2D ${\cal{PT}}$-symmetric photonic crystals we found that the frequency bands enter the region of broken ${\cal{PT}}$-symmetry
at the points of high symmetry $M$, $K$ and $\Gamma$ and remain in broken ${\cal{PT}}$-symmetric region as the in-plane wave vector is swept along high
symmetry directions. Specifically, bands 1 and 2 with real eigenvalues merge into a complex conjugate pair at the $M$-point and remain in broken ${\cal{PT}}$-symmetric
phase along $MK$ direction. Such a behavior is consistent with the general observation that the broken ${\cal{PT}}$-symmetry phase tends to occur at the Brillouin zone boundary \cite{ge}.
An interesting ${\cal{PT}}$-symmetry breaking occurs along $\Gamma K$ direction where band 2 merge subsequently with band 1 and band 3 at two adjacent exceptional points to form
the complex conjugate pair near the $K$-point and along $\Gamma K$ direction. In this case ${\cal{PT}}$-symmetric phase originates from the fact that both modes of
each pair possess the same symmetry along the direction orthogonal to the propagation direction. We note that ${\cal{PT}}$-symmetry breaking can occur also away of the Brillouin zone boundary, for example between bands 8 and 9 along $MK$ direction.

We claim that the resonant behavior of the normalized energy $S$ at the frequency $ f = 0.351$ can be assigned to the existence of complex conjugate pair at the $K$-point: The enhancement of the field may occur due to the negative imaginary part of the frequency associated
with the complex eigenmode corresponding to band 2 at this point -- see Fig.\ref{band-imag}. Simultaneously, the intensity of this mode can grow due to vanishing group velocity associated with the 2D standing mode at the $K$-point. Such a wave arises from the 2D diffraction which induces the coupling between the waves propagating along the three equivalent $\Gamma K$ directions along which the Bragg conditions are satisfied.
The existence of the standing wave pattern is consistent with the scattering diagram shown in Fig. \ref{S-1} which fully symmetric and reflects the arrangement of the identical cylinders with gain and loss along the $\Gamma$K directions. We note that similar standing wave associated with the second lowest resonance at $f = 0.58756$ arises from the coupling between the waves propagating along the three equivalent $\Gamma$M directions. In this case, however, the induced mode reflects $\cal{PT}$ symmetric arrangement of the gain/loss cylinders in 2D photonic lattice and thus in accord with asymmetric scattering response of PT dipole\cite{smk} exhibits  a strongly asymmetric diagram.

The resonant behavior of the normalized energy associated with the second lowest resonance cannot be assigned due to inherent limitations of our approach which does not allow to deal with the multiple couplings which occur at the $\Gamma$-point, namely between bands 2 and 3 at the frequency $f = 0.625$ and at the frequency $f = 0.64$ where two doubly degenerate pairs between bands 4,5 and 6,7 form four-fold degenerate supermode.



\section{Size effects in a finite 2D honeycomb lattice}

When the complex conjugate pair occurs at the edge of the first Brillouin zone,  where the group velocity of the wave is small,
the intensity of eigenfunction with negative imaginary part of the frequency starts to increase.
The amplification increases exponentially with time, $\propto\exp {\rm Im}~\omega t/(2\pi) $. Consequently, we expect that
larger samples radiate much stronger than the smaller.
In the first approximation, the intensity given by Eq. (2) can be expressed as
\be\label{eq:exp}
S = S_0e^{{\rm Im}~f~ t} = S_0e^{{\rm Im}~f~ L_s/v}
\ee
where $L_s$ is a size of the structure and $v=L_s/t$ is a group velocity. Consequently, scattered intensity should increase exponentially when the size of the sample increases.

\begin{figure}[t!]
\bc
\includegraphics[width=0.22\textwidth]{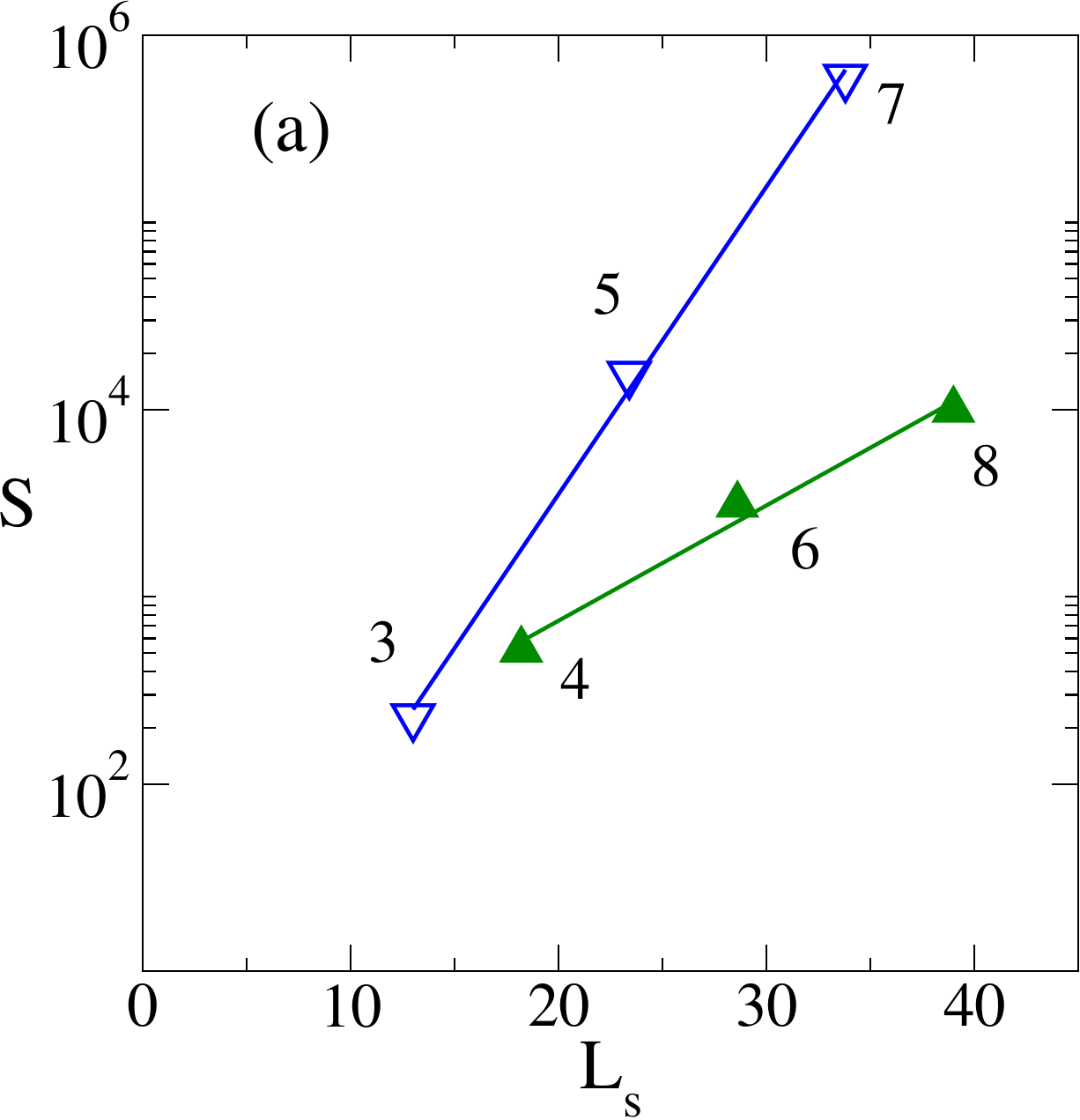}
~~
\includegraphics[width=0.22\textwidth]{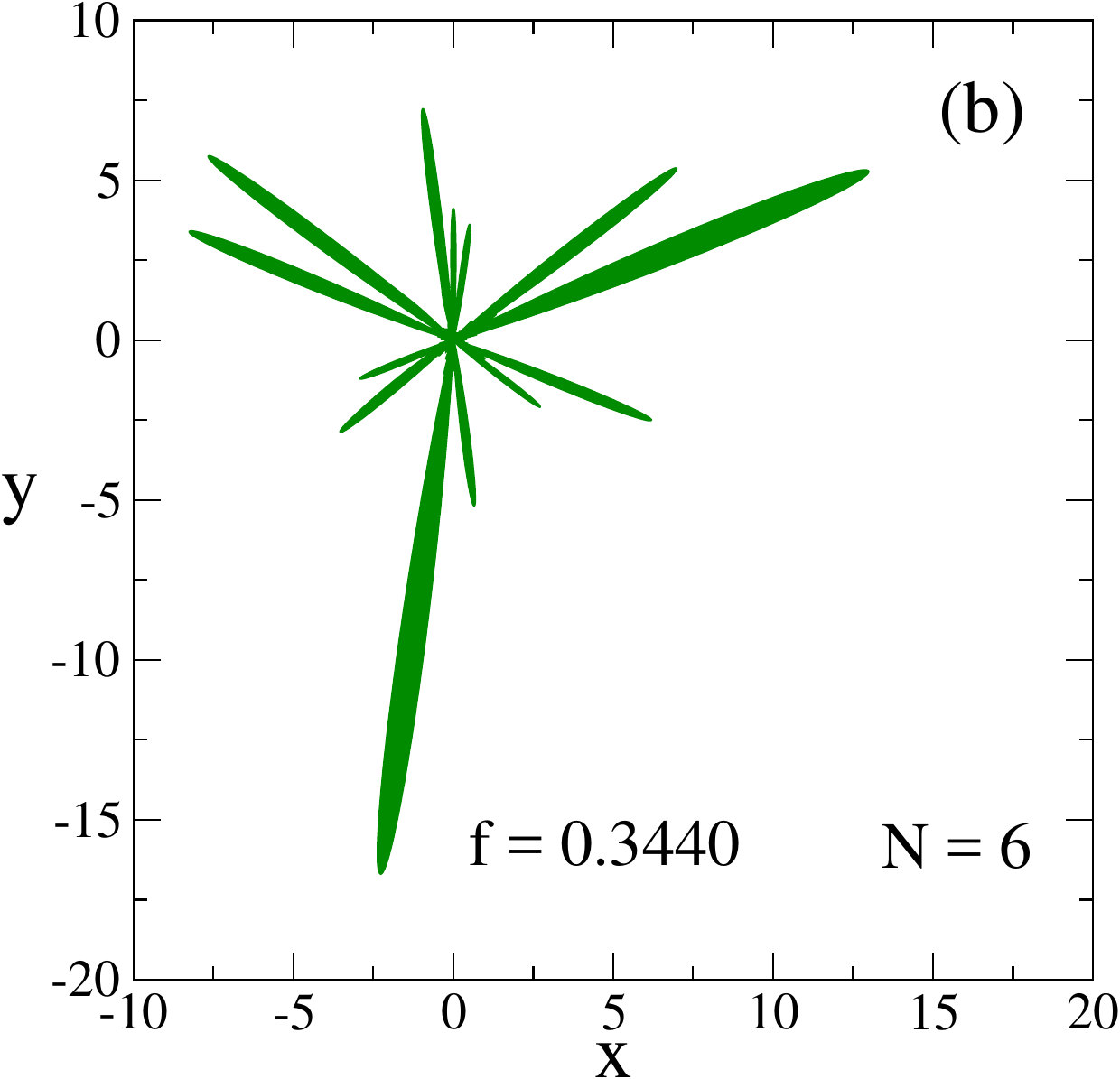}\\
\includegraphics[width=0.22\textwidth]{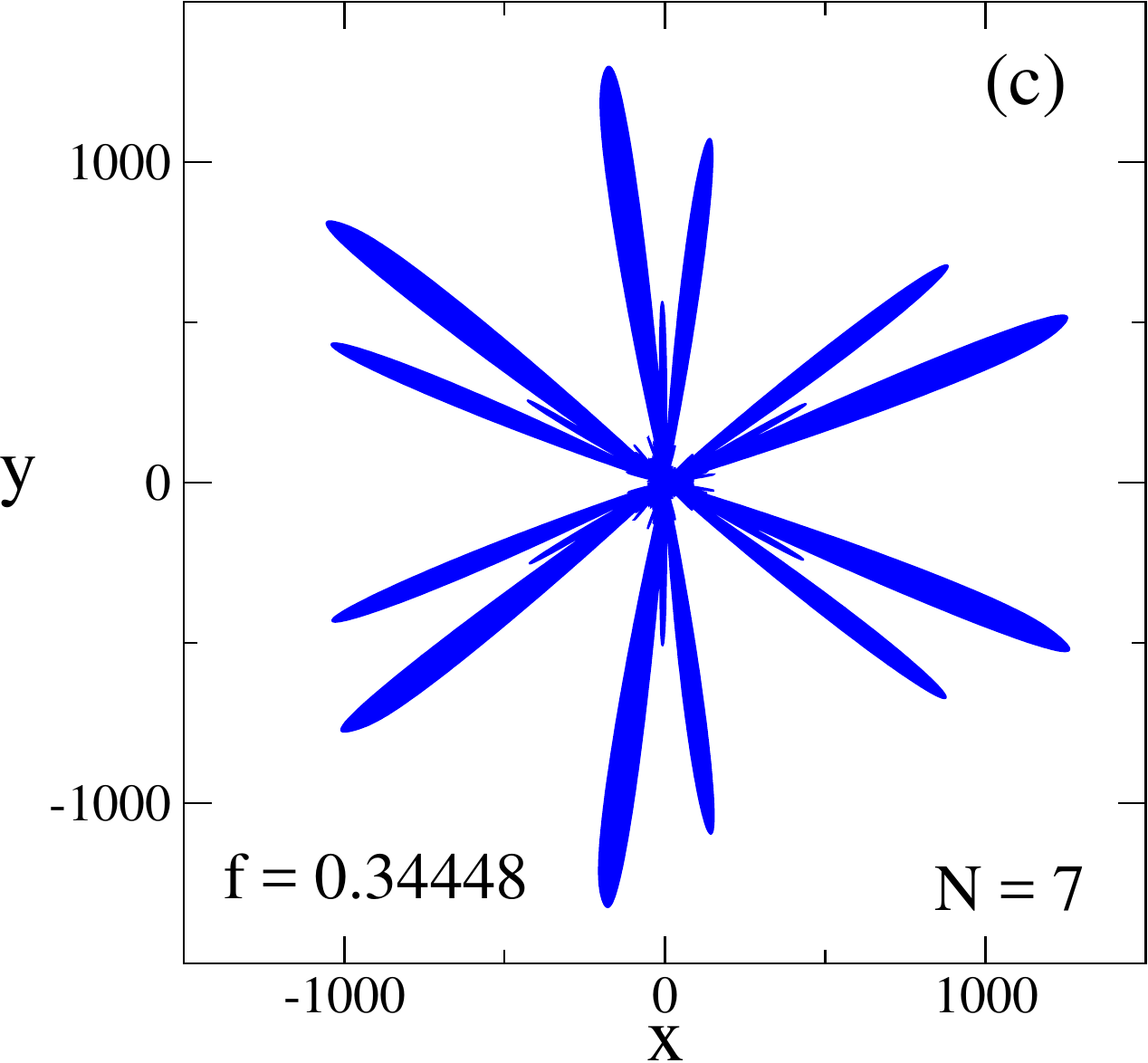}
~~
\includegraphics[width=0.22\textwidth]{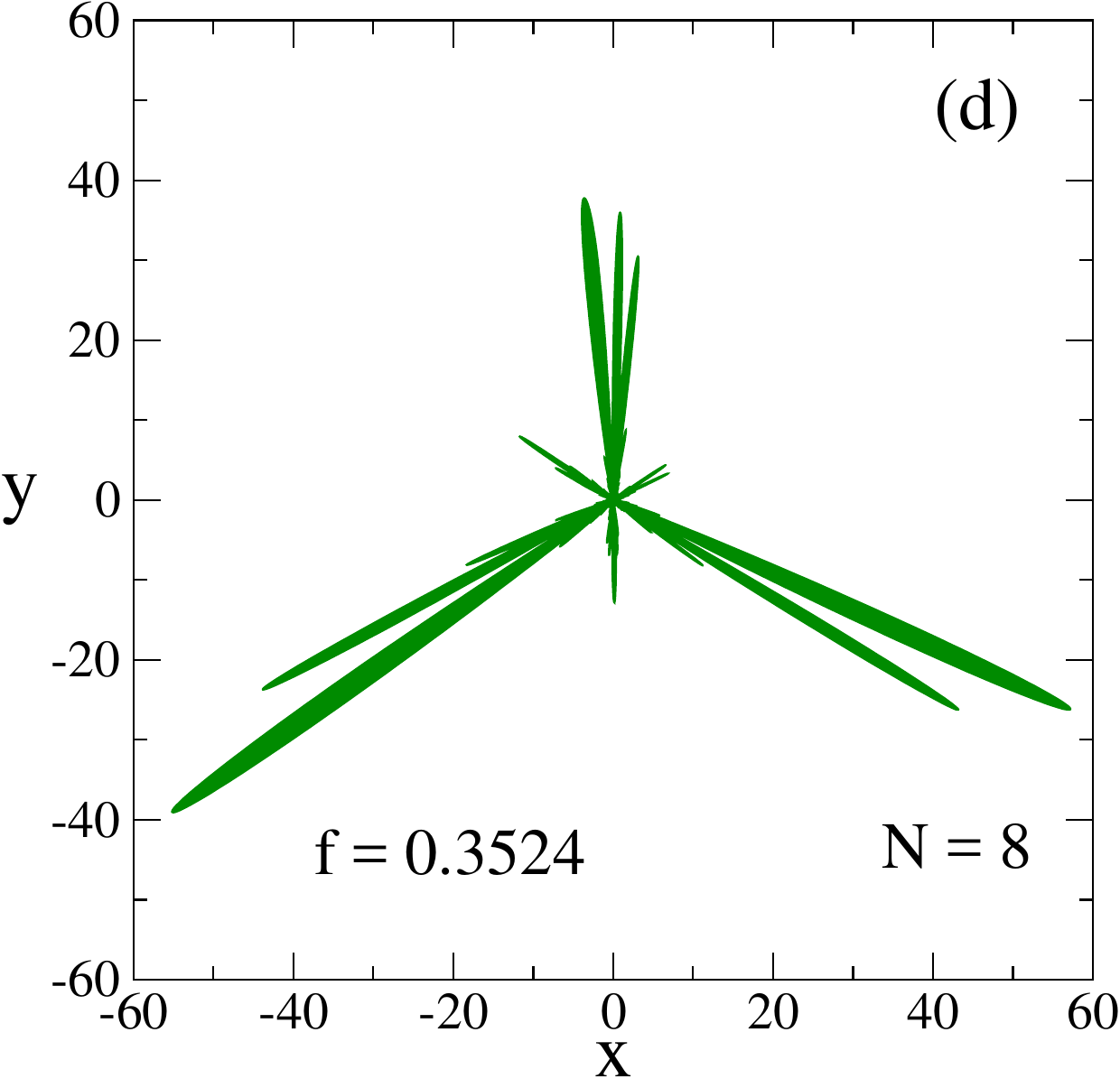}
\ec
\caption{(Color online)
(a)The intensity of the radiated power $S$, given by Eq. (2)
\textsl{vs} size of the sample $L_s$. The lattices associated with even ($N = 2, 4, 6$) and odd ($N = 3, 5, 7$) number of the hexagons in the
sample  basis are indicated by empty and full triangles, respectively. Three  other panels (b-d) show the scattering diagrams for $N= 6, 7$ and 8 ($R=300$). (The scattering for $N=5$ is shown in Fig. \ref{S-1a}(a)). Note different scales of the last three panels.}
\label{hexa3}
\end{figure}

To verify the exponential increase of the scattered intensity, we calculated scattering intensity $S$ for various size of the structure. As shown in Fig.~\ref{hexa3}(a), the intensity indeed increases exponentially with the size of the system, however, it differs considerably for odd and even $N$. As shown in Figs. \ref{hexa3}(b-d), not only size dependence, but also the scattering diagram depend on the parity of $N$. The features shown in Fig. \ref{hexa3} demonstrate that the optical properties of a finite size photonic crystal may deviate significantly from the periodic one at frequencies close to the band edge. The major difference stems from the fact that the finite structure possesses discrete spectrum of eigenstates instead of a continuum of modes. It has been shown in the case of square lattice of dielectric rods in air\cite{ruda}, that the eigenfrequencies of the confined modes which satisfy the conditions of constructive interference which depend on the positive integers $m$ and $n$ and on the size of the structure $L_s$.  Within this theoretical model based on envelope function approximation one can explain the dependence of the frequency of the resonances on the size of the sample as well as difference in the intensity of the radiated power as a function of the $L_s$ for even and odd $N$. The latter behavior can be assigned to the fact that the finite hexagonal structures with even and odd $N$ can be casted into two classes according to the symmetry along the $x$-axis and to the interference patterns that satisfy the corresponding boundary conditions.


\bc
\ec

\section{Conclusion}

We have shown numerically that a finite array  of ${\cal{PT}}$ dipoles consisting of gain/loss cylinders arranged in a 2D honeycomb lattice exhibits
profoundly different scattering properties in comparison with those consisting of lossless dielectric cylinders. In the far field limit, the structure scatters an incident electromagnetic plane wave with arbitrary frequency only in a few directions given by the spatial symmetry of the structure. In particular, we found that its total energy reveals series of sharp resonances at which at which the energy increases by two orders of magnitude.

Both features can be explained qualitatively in terms of the complex photonic band structure of an infinite 2D honeycomb lattice which supports broken ${\cal{PT}}$-symmetric phase at the high symmetry points and along the $\Gamma$K and $\Gamma$M directions and provides mechanisms leading to a significant enhancement of the radiated power and offering a plausible explanation to highly-directional scattering pattern. Specifically, we have assigned the lowest resonance in the total scattering energy at the frequency $f = 0.351$ to the broken ${\cal{PT}}$-symmetry mode formed by a doubly degenerate pair with complex conjugate eigenfrequencies at the $K$-point of the reciprocal lattice.
We have shown that the intensity of the radiated power at the resonance increases exponentially with the coefficient proportional to the imaginary part of the eigenfrequency and to the size of the sample and depends on the parity of the number of dipoles in the hexagon basis. A highly-directional scattering arises from the coupling between the waves propagating along the highly symmetric directions and its distribution reflects symmetry of the 2D photonic lattice along the relevant equivalent directions.

\section*{Acknowledgements}
The research of P.~Marko\v s was supported  by the Slovak Research and Development Agency under the contract No. APVV-15-0496
and by the Agency  VEGA under the contract No. 1/0108/17. The research of V. Kuzmiak was supported by
Grant 16-00329S of the Czech Science Foundation(CSF).

\end{document}